\documentclass{article}

\usepackage{amsmath}
\usepackage{amsfonts}
\usepackage{graphics}
\usepackage{pstricks}
\usepackage{array}
\usepackage{shortvrb}
\usepackage{epsf}
\usepackage{graphicx}
\usepackage{rotating}
\usepackage{multirow}
\usepackage{cite}
\usepackage{flushend}
\usepackage{color}
\usepackage{alltt}
\usepackage{colortbl}
\usepackage{lineno}
\usepackage{soul}
\usepackage{amssymb}

\newcommand{\Ito}{It\^{o}}

\begin{document}

 \title{Construction of SDE-based wind speed models with exponential 
autocorrelation}
 \author{Rafael Z\'{a}rate-Mi\~{n}ano, Federico Milano}
 \date{September 2015}
 \maketitle

\begin{abstract}
This paper provides a systematic method to build wind speed models 
based on stochastic differential equations (SDEs). The resulting models produce 
stochastic processes with a given probability distribution and exponential 
decaying autocorrelation function. The only information needed to build the 
models is the probability density function of the wind speed and its 
autocorrelation coefficient. Unlike other methods previously proposed in the 
literature, the proposed method leads to models able to reproduce an exact 
exponential autocorrelation even if the probability distribution is not 
Gaussian. A sufficient condition for the property above is provided.
The paper includes the explicit formulation of SDE-based wind 
speed models obtained from several probability distributions used in the 
literature to describe different wind speed behaviors.
\end{abstract}

\section{Introduction} \label{sec:intro}
Wind speed models are used in the analysis of many aspects related to power 
systems, for example, in power system economics and operation (e.g., 
\cite{OlssonSoder:2010, MoralesPerezRuiz:2011, PapavasiliouONeill:2011}), 
generation capacity reliability evaluation (e.g.,
\cite{BillintonBai:2004, LeiteFalcao:2006, BillintonHuang:2011}), and dynamic 
studies and control of wind turbines (e.g., \cite{NichitaCeanga:2002, 
MuhandoFunabashi:2008, AndrewZhe:2010, MelicioCatalao:2011}). The types of 
models traditionally used in the different research fields include time series, 
four-component composite models, and models based on Kalman filters. 
Independently of the type of the model, the appropriate characterization of 
the wind behaviour is a key modeling aspect, since the reliability of the 
results obtained in the above studies depends on it. In this 
paper, we develop a novel method based on stochastic differential equations, 
the regression theorem, and the Fokker-Planck equation, to construct wind speed 
models.

From a statistical point of view, the wind speed is characterized 
by its probability distribution and autocorrelation. Therefore, to be adequate, 
wind speed models should be able to reproduce such characteristics. The type of 
probability distribution that best describes the wind variability depends on the 
particular location and on the time frame \cite{CartaVelazquez:2009, 
CalifSoubdhan:2011, LoBranoCulotta:2011, Calif:2012}. With regard to the 
autocorrelation of the wind speed, this has been
usually characterized by an exponential decaying function, either for hourly 
wind speed measurements in the time frame of hours \cite{BrettTuller:1991}, or 
for wind speed measurements on a one-second basis in the time frame of minutes 
\cite{Calif:2012}. However, other studies have identified scaling properties 
in the wind speed measurements at different sites where the autocorrelation is 
better described by means of power-law decaying functions 
\cite{KavasseriNagarajan:2005, CalifSchmitt:2014}. This paper focuses on 
the 
development of wind speed models for locations where the autocorrelation 
observed in the wind speed is of exponential type. Therefore, the validity 
of the proposed models is limited to cases for which such a condition is 
satisfied.

The application of stochastic differential equations (SDEs) 
to the modeling of stochastic processes occurring in power systems is gaining 
interest in recent years (e.g, \cite{DongHill:2012, MilanoZarate:2013, 
DhopleDominguezGarcia:2013}). A SDE is composed of two terms: the drift term and 
the diffusion term. The specific formulation of each term determines the 
statistical properties of the phenomenon under consideration. With this 
regard, SDEs have been successfully applied to wind speed fluctuation modeling 
when such fluctuations show an exponentially autocorrelated Gaussian behaviour 
\cite{Calif:2012}.
However, the construction of SDEs to model exponentially 
autocorrelated non-gaussian phenomena, as it can be the case of hourly wind 
speeds, is still an open task. 

In a previous work, \cite{ZarateMilano:2013}, we proposed to 
overcome this difficulty by transforming a well-known SDE widely used to model 
exponentially autocorrelated Gaussian processes. For that, translation 
techniques are applied in order to obtain another SDE that reproduce a given 
non-gaussian probability distribution. The resulting model is able to 
reproduce such probability distribution but it cannot guarantee a good 
reproduction of the autocorrelation of the process.

The method proposed in this paper relies on basic stochastic calculus 
concepts (such as the Regression Theorem) to derive an expression for the drift 
term of the SDE that ensures an exponentially autocorrelated process. Then, the 
stationary Fokker-Planck equation is solved to obtain the expression of the 
diffusion term that guarantee a given probability distribution. Therefore, 
the 
models that result from applying the proposed method are able to exactly 
reproduce both the probability distribution and the exponential autocorrelation 
for which they are designed.

The proposed method is systematically applied to construct SDE-based models 
from different probability distributions proposed in the literature to describe 
the wind speed behaviour. As a result, together with the detailed description 
and justification of the proposed method, the paper provides a collection 
of 
SDE-based models ready to be used in different studies related to wind power. 
Although the development of the method is motivated by wind speed 
modeling, the proposed technique is general, and it can applied to model 
phenomena other than wind speed.

The remainder of the paper is organized as follows. Section
\ref{sec:theory} provides the main theoretical background required to 
developing the proposed method. Section \ref{sec:devel} describes and 
justifies the procedure that leads to the mathematical formulation of the wind 
speed models. Examples of SDEs that generate exponentially autocorrelated 
stochastic process for several different probability distribution functions are 
given in Section \ref{sec:examples}. Finally, Section \ref{sec:con} provides 
relevant conclusions.

\section{Outlines of Stochastic Calculus}
\label{sec:theory}
This section outlines briefly the relevant concepts of stochastic calculus 
that are required to define the proposed wind speed models. The interested 
reader find more details, for example in \cite{Oksendal:2003, Grigoriu:2002, 
Gardiner:2009}. For simplicity, but without loss of generality, all results are 
shown for one-dimensional processes.

\subsection{Stationary Markov processes}
\label{sec:sp}
The following two definitions are taken from \cite{Grigoriu:2002}:

\begin{itemize}

\item D1 (Markov processes). A stochastic process $x(t)$ is Markov if for 
every integer $n \geq 1$ and times $t_1 < \cdots < t_n$, the conditional random 
variables $x(t_n)|x(t_{n-1})$ and $(x(t_{n-2}), \cdots, x(t_{n-1}))|x(t_{n-1})$ 
are independent, that is, the past of $x(t)$ is independent on its future 
conditional on the present.

\item D2 (Stationary processes). A stochastic process is said to be stationary, 
or stationary in the strict sense, if for every integer $n \geq 1$, times $t_1 
< \cdots < t_n$, and time shift $\tau$, $(x(t_1), \cdots, x(t_n)) 
\stackrel{d}{=} (x(t_{1}+\tau), \cdots, x(t_{n}+\tau))$, that is, the 
probability distribution of $x(t)$ is time invariant. 
\end{itemize}

Therefore, if a process $x(t)$ is stationary, its mean and standard 
deviation are time invariant, that is, $E[x(t)] = \mu(t) = \mu$ and 
$\sqrt{E[(x(t)-\mu)^2]} = \sigma(t) = \sigma$, and the autocovariance $c(s, 
t)$ and autocorrelation $r(s, t)$ functions

\begin{align}
 &c(s, t) = E\left[(x(s)-\mu(s)) \cdot (x(t)-\mu(t))\right] \\
 \nonumber \\
 &r(s, t) = 
\frac{E\left[(x(s)-\mu(s)) \cdot (x(t)-\mu(t))\right]}{\sigma(s) \cdot 
\sigma(t)}
\end{align}

depend only on the time lag $\tau = t - s$, that is

\begin{align}
 &c(s, t) = c(\tau) = E\left[(x(t-\tau)-\mu) \cdot (x(t)-\mu)\right] \\
 \nonumber \\
 &r(s, t) = r(\tau) = 
\frac{E\left[(x(t-\tau)-\mu) \cdot (x(t)-\mu)\right]}{\sigma^2}
\end{align}

where $E[\cdot]$ is the expectation operator and $\sigma^2$ is the variance of 
the stationary stochastic process.

\subsection{Regression theorem} \label{sec:Regre}
In the theory of stochastic processes, the regression theorem states that for 
Markov processes in which the mean values obey linear evolution equations of 
the form

\begin{equation} \label{eq:RegreMean}
\frac{dE[x(t)]}{dt} = -\alpha \cdot E[x(t)]
\end{equation}

then, in the stationary state, the stationary autocovariance function 
$c(\tau)$ can be obtained by solving

\begin{equation} \label{eq:RegreCov}
\frac{dc(\tau)}{d\tau} = -\alpha \cdot c(\tau)
\end{equation}

with initial condition $c(0) = \sigma^2$ \cite{Gardiner:2009}. The result of 
solving (\ref{eq:RegreCov}) is

\begin{equation} \label{eq:Cov}
c(\tau) = \sigma^2 \cdot e^{-\alpha \cdot \tau}
\end{equation}

showing that the autocovariance function of such processes is an exponential 
decaying function. Similarly, and assuming an initial condition 
$r(0) = 1$, for the autocorrelation we obtain

\begin{equation} \label{eq:Aut}
r(\tau) = e^{-\alpha \cdot \tau}
\end{equation}

which is also an exponential decaying function.

\subsection{\Ito{} stochastic differential equations}
\label{sec:SDE}
A one-dimensional \Ito{} Stochastic Differential Equation (SDE) has the 
general form

\begin{align} \label{eq:sde}
  &dx(t) = a(x(t), t) \cdot dt + b(x(t), t) \cdot dW(t), \quad t \in [0, T], \\ 
\nonumber
  &x(0) = x_0, 
\end{align}

where the initial value $x_0$ can be a deterministic or a random value, and 
$W(t)$ is a standard Wiener process, also loosely called Brownian motion 
\cite{Gardiner:2009, Grigoriu:2002}. The integral form of equation 
(\ref{eq:sde}) is

\begin{equation} \label{eq:sde_int}
  x(t) - x_{0} = \int_{0}^{t} a(x(u), u) \cdot du + \int_{0}^{t} b(x(u), 
s) \cdot dW(u), 
\quad t \in [0, T], 
\end{equation}

where the first integral is an ordinary Riemann-Stieltjes integral and the
second one is a stochastic integral interpreted in the \Ito{}'s sense. The 
solution of (\ref{eq:sde}) or (\ref{eq:sde_int}) is a stochastic 
process so-called diffusion process, and functions $a(x(t),t)$ and $b(x(t),t)$ 
are referred to as the drift and the diffusion terms of the \Ito{} SDE, 
respectively. Diffusion processes are continuous-time Markov processes with 
almost surely continuous sample paths \cite{Grigoriu:2002}.

\subsection{The \Ito{} formula} \label{sec:Ito}
The \Ito{} formula can be viewed as the stochastic equivalent of the 
deterministic chain rule and allows finding the differential of a function 
whose arguments include a stochastic variable. If the stochastic 
variable is defined by a SDE of the type of (\ref{eq:sde}) the \Ito{} formula 
gives the SDE of the same type that the function obeys. For an arbitrary 
function $g(\cdot)$ of the stochastic variable $x(t)$ defined by 
(\ref{eq:sde}), the differential of $g(\cdot)$ is

\begin{align} \label{eq:Ito_dif}
  dg&(x(t),t) = \nonumber \\ 
  &\left[\frac{\partial g(x(t), t)}{\partial t} 
+ a(x(t), t) \cdot \frac{\partial g(x(t), t)}{\partial x(t)}
	  + \frac{1}{2} \cdot b^{2}(x(t), t) \cdot \frac{\partial^2 g(x(t), 
t)}{\partial x^2(t)}\right] \cdot dt \nonumber \\
  & + b(x(t), t) \cdot \frac{\partial g(x(t), t)}{\partial x(t)} \cdot dW(t)
\end{align}

where $a(x(t), t)$ and $b(x(t), t)$ are the drift and the 
diffusion terms of (\ref{eq:sde}), respectively \cite{Grigoriu:2002, 
Gardiner:2009}.

The \Ito{} formula can be used to develop differential equations, for 
example, for moments and probability densities of the stochastic process $x(t)$.

\subsection{Fokker-Planck equation} \label{sec:FPE}
The Fokker-Planck equation describes the time evolution of the probability 
densities of a stochastic process. For the stochastic process $x(t)$ defined by 
the SDE (\ref{eq:sde}), the associated Fokker-Planck equation has the following 
form:

\begin{equation}
 \label{eq:FPE}
  \frac{\partial p(x(t), t)}{\partial t} = - \frac{\partial}{\partial x(t)} 
\left[a(x(t), t) \cdot p(x(t), t) \right] + \frac{1}{2} \cdot 
\frac{\partial^2}{\partial x^2(t)} \left[b^2(x(t), t) \cdot p(x(t), t) \right]
\end{equation}

where functions $a(x(t), t)$ and $b(x(t), t)$ in (\ref{eq:FPE}) are the drift 
and diffusion terms of the SDE (\ref{eq:sde}), respectively, and $p(x(t), t)$ 
is the transient probability density function of the process $x(t)$. In 
(\ref{eq:FPE}) it is assumed that the following conditions are satisfied 
\cite{Grigoriu:2002}:

\begin{equation}
 \label{eq:FPE_cond1}
  \lim_{|x(t)|\rightarrow \infty}\left[a(x(t), t) \cdot p(x(t), t) \right] = 0
\end{equation}
\begin{equation}
 \label{eq:FPE_cond2}
  \lim_{|x(t)|\rightarrow \infty}\left[b^2(x(t), t) \cdot p(x(t), t) \right] = 0
\end{equation}
\begin{equation}
 \label{eq:FPE_cond3}
  \lim_{|x(t)|\rightarrow \infty}\left[\frac{\partial}{\partial 
x(t)} \left[b^2(x(t), t) \cdot p(x(t), t) \right] \right] = 0
\end{equation}

\section{Proposed Building Method of the SDE Model}
\label{sec:devel}
In this section, we develop the proposed method on the basis of the stochastic 
calculus results exposed in the previous section.

Our goal is to build a SDE model to generate an exponentially 
autocorrelated stochastic process with a given probability distribution. In 
other words, we look for the form of the drift and diffusion terms of equation 
(\ref{eq:sde}) so that the solution of the resulting SDE is a process with 
those statistical properties.

Inspired in the approach of \cite{Calif:2012}, our method is based on the 
relation that the drift and the diffusion terms should hold in order to get a 
given probability distribution. This relation is obtained from the 
Fokker-Planck 
equation. For stationary process, $a(x(t), t) = a(x(t))$,  $b(x(t), t) = 
b(x(t))$, and $p(x(t), t) = p(x(t))$, and the Fokker-Planck equation 
(\ref{eq:FPE}) reduces to

\begin{align} \label{eq:FPE_st1}
  0 = - \frac{\partial}{\partial x(t)} 
\left[a(x(t) \cdot p(x(t)) \right] + \frac{1}{2} 
\cdot \frac{\partial^2}{\partial x^2(t)} 
\left[b^2(x(t)) \cdot p(x(t)) \right]
\end{align}

Therefore, functions $a(x(t))$, $b(x(t))$, and $p(x(t))$ are related by means 
of the following equation

\begin{align} \label{eq:FPE_st2}
  0 = - a(x(t)) \cdot p(x(t)) + \frac{1}{2} \cdot \frac{\partial}{\partial 
x(t)}\left[b^2(x(t)) \cdot p(x(t)) \right] + C
\end{align}

where constant $C$ is zero to satisfy conditions 
(\ref{eq:FPE_cond1})-(\ref{eq:FPE_cond3}).

By solving (\ref{eq:FPE_st2}) for $a(x(t))$ we obtain

\begin{align} \label{eq:cond_a}
  a(x(t)) = b(x(t)) \cdot \frac{\partial b(x(t))}{\partial x(t)} + 
\frac{1}{2} \cdot b^2(x(t)) \cdot \frac{\partial \ln{p(x(t))}}{\partial x(t)} 
\end{align}

and, by solving (\ref{eq:FPE_st2}) for $b^2(x(t))$ we obtain

\begin{align} \label{eq:cond_b}
  b^2(x(t)) = \frac{2}{p(x(t))} \cdot \int^{x(t)}_{-\infty}{a(z(t)) \cdot 
p(z(t))} \cdot dz(t)
\end{align}

for $p(x(t)) \neq 0$, and $b(x(t)) = 0$ if $p(x(t)) = 0$. Therefore, for a 
given probability density function $p(x(t))$, if one of the functions $b(x(t))$ 
or $a(x(t))$ is known, the other function can be obtained by solving 
(\ref{eq:cond_a}) or (\ref{eq:cond_b}), respectively.

In reference \cite{Calif:2012} the diffusion term $b(x(t))$ is fixed to a 
constant value according to Kolmogorov theory of local isotropy 
\cite{Pope:2000}, and 
the drift term $a(x(t))$ is obtained by solving (\ref{eq:cond_a}) for different 
probability distributions. With this approach, the resulting SDE provides a 
stochastic process with the given probability distribution, but the 
empirical exponential decay of the autocorrelation is not properly 
reproduced for non-gaussian processes. We proceed in a different way: first, we 
obtain a drift term $a(x(t))$ that ensures an exponential autocorrelation 
function with a given decay rate. Second, we obtain the diffusion term 
$b(x(t))$ by solving (\ref{eq:cond_b}) for the given probability density 
function $p(x(t))$.

To identify the desired drift function, we apply the \Ito{} formula 
(\ref{eq:Ito_dif}) to develop a differential equation of the 
stationary autocovariance of a process modeled with (\ref{eq:sde}). For that, 
function $g(\cdot)$ is selected to be 

\begin{equation}
g(x(t)) = (x(s) - \mu) \cdot (x(t)-\mu)
\end{equation}

where $s < t$. The derivatives involved in (\ref{eq:Ito_dif}) are as follows:

\begin{align}
&\frac{\partial g(x(t))}{\partial t} = 0 \label{eq:dg1}\\
&\frac{\partial g(x(t))}{\partial x(t)} = x(s) - \mu \\
&\frac{\partial^2 g(x(t))}{\partial x^2(t)} = 0 \label{eq:dg3}
\end{align}

Observe that, in the previous derivations, we have used the fact that $x(s)$ is 
independent of $x(t)$ due to the Markov property, and that the chosen function 
$g(x(t))$ does not explicitly depend on time. From (\ref{eq:Ito_dif}) and 
(\ref{eq:dg1})-(\ref{eq:dg3}), the resulting SDE is

\begin{align} \label{eq:g_dif}
  d [ (x(s) -& \mu) \cdot (x(t)-\mu)]  = \nonumber \\ 
  &a(x(t)) \cdot (x(s) - \mu) \cdot dt + b(x(t)) \cdot (x(s) - \mu) 
\cdot dW(t)
\end{align}

with initial condition $(x(s)-\mu)^2$. The integral form of the previous SDE 
is

\begin{align} \label{eq:g_int}
  (&x(s) - \mu) \cdot (x(t)-\mu) - (x(s)-\mu)^2 = \nonumber \\ 
  &\int_{s}^{t} a(x(u)) \cdot (x(s) - \mu) \cdot du 
  + \int_{s}^{t} b(x(u)) \cdot (x(s) - \mu) \cdot dW(u)
\end{align}

where we perform the integration over the interval $[s, t]$. By applying the 
expectation operator $E[\cdot]$ to equation (\ref{eq:g_int}), and taking into 
account that the expectation of an \Ito{} stochastic integral is zero 
\cite{Oksendal:2003}, i.e.,

\begin{equation}
  E \left[ \int_{s}^{t} b(x(u)) \cdot (x(s) - \mu) \cdot dW(u) \right] = 0
\end{equation}

we obtain the following expression

\begin{align} \label{eq:cov_int}
  E\left[(x(s) - \mu) \cdot (x(t)-\mu)\right] - 
&E\left[(x(s)-\mu)^2\right] 
= \nonumber \\ 
  &\int_{s}^{t} E\left[a(x(u)) \cdot (x(s) - \mu) \right] \cdot du
\end{align}

where the first term of the right hand side of equation (\ref{eq:cov_int}) is 
the autocovariance function. The differential form of (\ref{eq:cov_int}) is

\begin{equation} \label{eq:cov_dif1}
  \frac{dE\left[(x(s) - \mu) \cdot (x(t)-\mu)\right]}{dt} = 
  E\left[a(x(t)) \cdot (x(s) - \mu)\right]
\end{equation}

In order to obtain an equation similar to (\ref{eq:RegreCov}) it is clear that

\begin{equation} \label{eq:drift1}
  a(x(t)) = -\alpha \cdot (x(t) - \mu)
\end{equation}

and (\ref{eq:cov_dif1}) can be expressed as

\begin{equation} \label{eq:cov_dif2}
  \frac{dc(s, t)}{dt} = -\alpha \cdot c(s, t)
\end{equation}

For stationary processes, the autocovariance only depends on the time lag 
$\tau = t-s$, therefore equation (\ref{eq:cov_dif2}) reduces to 
(\ref{eq:RegreCov}), and the autocovariance and the autocorrelation of the 
stochastic process $x(t)$ follow the decaying exponential expressions 
(\ref{eq:Cov}) and (\ref{eq:Aut}), respectively. 

Observe also that as the drift term (\ref{eq:drift1}) is linear, the 
requirement of a linear evolution equation for the mean value expressed in the 
regression theorem is also satisfied. This can be shown from the integral 
version of a generic SDE with the computed drift term, i.e.,

\begin{equation} \label{eq:sde_int2}
  x(t) - x_{0} = \int_{0}^{t} -\alpha \cdot (x(u) - \mu) \cdot du + 
\int_{0}^{t} b(x(u)) \cdot dW(u)
\end{equation}

By applying the expectation operator to equation (\ref{eq:sde_int2}), and 
taking into account that the expectation of an \Ito{} stochastic integral is 
zero, we obtain

\begin{equation} \label{eq:Esde_int}
  E[x(t)] - E[x_{0}] = \int_{0}^{t} -\alpha \cdot E \left[ (x(u) - \mu) \right] 
\cdot du
\end{equation}

and, recovering the differential form,

\begin{equation} \label{eq:Mean_dif}
\frac{dE[x(t)]}{dt} = -\alpha \cdot E[x(t)] + \alpha \cdot \mu
\end{equation}

with initial condition $E[x_0]$. Observe that equation (\ref{eq:Mean_dif}) 
expresses a linear law similar to (\ref{eq:RegreMean}).

In summary, to model a stationary stochastic process with given 
probability distribution function $p(x(t))$ and exponential autocorrelation 
with a SDE, it is a sufficient condition to define a drift term in the form

\begin{equation}\label{eq:drift2}
  a(x(t)) = -\alpha \cdot (x(t) - \mu)
\end{equation}

where $\mu$ is the mean of the particular probability distribution $p(x(t))$, 
and a diffusion term computed by solving

\begin{equation} \label{eq:diff}
  b^2(x(t)) = \frac{2}{p(x(t))} \int^{x(t)}_{-\infty}{-\alpha \cdot (z(t)-\mu) 
\cdot p(z(t))} \cdot dz(t)
\end{equation}

\section{Examples}
\label{sec:examples}
In this section, we apply the proposed method to construct SDE-based wind speed 
models for different probability distributions that have been proposed in the 
literature to describe the wind speed variability. In Subsections 
\ref{sec:normal} and \ref{sec:gc3o} we use the Normal 
distribution and the Gram-Charlier expansion proposed in \cite{Calif:2012}, 
respectively to fit wind speed fluctuations around a mean value measured on a 
one-second basis. 
In Subsections \ref{sec:b3p}-\ref{sec:w2p} we use a variety of probability 
distributions analyzed in \cite{CartaVelazquez:2009} to fit hourly mean wind 
speeds recorded at different weather stations. To simplify the notation, 
the explicit dependency of variable $x$ on time is removed. All models have the 
following structure:

\begin{equation}
 dx = a(x) \cdot dt + b(x) \cdot dW(t)
\end{equation}

where $a(x)$ and $b(x)$ are defined according to (\ref{eq:drift2}) and 
(\ref{eq:diff}), respectively.

\subsection{Normal distribution}
\label{sec:normal}
The probability density function $p_{\rm N}(x)$ of the Normal distribution is

\begin{equation} \label{eq:pdf_N}
 p_{\rm N}(x) = 
\frac{1}{\sigma \cdot \sqrt{2 \cdot \pi}} \cdot 
\exp \left(\displaystyle - \frac{\left( x-\mu \right)^2}{2 \cdot \sigma^2} 
 \right)
\end{equation}

where $\mu$ is the mean, and $\sigma$ is the standard deviation.

By applying the proposed method, the drift term is 

\begin{equation} \label{eq:a_N}
a(x) = -\alpha \cdot \left( x - \mu \right),
\end{equation}

and the diffusion term is

\begin{equation}\label{eq:b_N}
b(x) = \sqrt{2 \cdot \alpha} \cdot \sigma
\end{equation}

\subsection{Gram-Charlier III-order expansion}
\label{sec:gc3o}
The Gram-Charlier expansions are generally used to describe deviations from the 
Normal distribution by means of the incorporation of the skewness and kurtosis 
factors to the distribution. In particular, the Gram-Charlier III-order 
expansion has the following probability density function: 

\begin{equation}
 p_{\rm GC}(x) = \left( 1 + 
\frac{S}{6} \cdot {\rm He_{3}} \left(\frac{x-\mu}{\sigma}\right)\right) \cdot 
p_{\rm N}(x)
\end{equation}

where $p_{\rm N}(x)$ is the Normal probability density function 
(\ref{eq:pdf_N}), $S$ is the skewness factor, and

\begin{equation}
{\rm He_{3}} \left(\frac{x-\mu}{\sigma}\right) = 
\left(\frac{x-\mu}{\sigma}\right)^3 - 
3 \, \left(\frac{x-\mu}{\sigma}\right)
\end{equation}

is the Hermite polynomial of order 3.

For the standard Normal distribution $N(0, 1)$ the probability density 
function $p_{\rm GC}(x)$ is

\begin{equation}
 p_{\rm GC}(x) = \left( 1 + 
\frac{S}{6} \cdot \left( x^3 - 3 \cdot x \right)\right) \cdot \frac{1}{\sqrt{2 
\cdot \pi}} \exp \left(-\frac{1}{2} \cdot x^2 \right)
\end{equation}

By applying the proposed method, the drift term is 

\begin{equation} \label{eq:a_gc3o}
a(x) = -\alpha \cdot x
\end{equation}

and the diffusion term is

\begin{equation} \label{eq:b_gc3o}
b(x) = \sqrt{\frac{2 \cdot \alpha \cdot \left( S \cdot x^3 + 6\right)}{S 
\cdot x \cdot \left( x^2 - 3\right) + 6}}
\end{equation}

\subsection{Three-parameter Beta distribution}
\label{sec:b3p}
The probability density function $p_{\rm B}(x)$ of the three-parameter Beta 
distribution is

\[
p_{\rm B}(x) =
\begin{cases}
\displaystyle
\frac{1}{\lambda_3 \cdot B \left(\lambda_1,\lambda_2 \right)} \cdot \left( 
\displaystyle\frac{x}{\lambda_3} \right)^{\lambda_1-1} \cdot \left( 
\displaystyle \frac{\lambda_3-x}{\lambda_3}  \right)^{\lambda_2-1} 
 & \text{if } x > 0 \\
0 & \text{if } x \leq 0
\end{cases}
\]

where $B(\cdot,\cdot)$ is the Beta function, $\lambda_1$ and 
$\lambda_2$ are shape parameters, and $\lambda_3$ is a noncentrality 
parameter.

By applying the proposed method, the drift term is

\begin{equation}\label{eq:a_b3p}
a(x) = -\alpha \cdot \left(\displaystyle x - \frac{\lambda_1 \cdot 
\lambda_3}{\lambda_1 + \lambda_2}\right)
\end{equation}

and the diffusion term is

\begin{equation}\label{eq:b_b3p}
b(x) = \sqrt{\frac{2 \cdot \alpha \cdot \left( \lambda_3 - x \right) \cdot 
x}{\lambda_1 + \lambda_2}}
\end{equation}

\subsection{Two-parameter Gamma distribution}
\label{sec:g2p}
The probability density function $p_{\rm G}(x)$ of the two-parameter Gamma 
distribution is

\[
p_{\rm G}(x) =
\begin{cases}
\displaystyle
\frac{1}{\displaystyle {\lambda_2}^{\lambda_1} \cdot
\Gamma \left(\lambda_1 \right)} \cdot x^{\lambda_1-1} \cdot \exp 
\left(\displaystyle 
-\frac{x}{\lambda_2} \right) & \text{if } x > 0 \\
0 & \text{if } x \leq 0
\end{cases}
\]

where $\Gamma(\cdot)$ is the Gamma function, $\lambda_1$ is a shape parameter, 
and $\lambda_2$ is a scale parameter.

By applying the proposed method, the drift term is

\begin{equation}\label{eq:a_g2p}
a(x) = -\alpha \cdot \left(x - \lambda_1 \cdot \lambda_2 \right)
\end{equation}

and the diffusion term is

\begin{equation}\label{eq:b_g2p}
b(x) = \sqrt{2 \cdot \alpha \cdot \lambda_2 \cdot x}
\end{equation}

\subsection{Three-parameter Generalized Gamma distribution}
\label{sec:g3p}
The probability density function $p_{\rm GG}(x)$ of the three-parameter 
Generalized Gamma distribution is

\[
p_{\rm GG}(x) =
\begin{cases}
\displaystyle
\frac{1}{\displaystyle {\lambda_2} \cdot \Gamma \left(\lambda_1 \right)} \cdot 
\lambda_3 \cdot \left(\displaystyle \frac{x}{\lambda_2}\right)^{\lambda_1 \cdot 
\lambda_3 -1} \cdot \exp \left( -\left( \displaystyle \frac{x}{\lambda_2} 
\right)^{\lambda_3} \right) & \text{if } x > 0
\\
0 & \text{if } x \leq 0
\end{cases}
\]

where $\Gamma(\cdot)$ is the Gamma function, $\lambda_1$ and $\lambda_3$ are 
shape parameters, and $\lambda_2$ is a scale parameter.

By applying the proposed method, the drift term is

\begin{equation}\label{eq:a_g3p}
a(x) = -\alpha \cdot \left(x - \displaystyle \frac{\lambda_2 \cdot \Gamma 
\left( \lambda_1 + \displaystyle 
\frac{1}{\lambda_3}\right)}{\Gamma(\lambda_1)} 
\right)
\end{equation}

and the diffusion term is

\begin{equation}
b(x) = \sqrt{b_1(x) \cdot b_2(x)}
\end{equation}

with

\begin{equation}
b_1(x) = 2 \cdot \alpha \cdot \lambda_2 \cdot x \cdot \left(\displaystyle 
\frac{x}{\lambda_2} \right)^{-\lambda_1 \cdot \lambda_3} \cdot \exp \left( 
\left(\displaystyle \frac{x}{\lambda_2} \right)^{\lambda_3} \right)
\end{equation}

and

\begin{equation}\label{eq:b_g3p}
b_2(x) = \frac{\Gamma \left( \lambda_1 \right) \cdot \Gamma \left(\displaystyle  
\lambda_1 + \frac{1}{\lambda_3}, \left( \frac{x}{\lambda_2} \right)^{\lambda_3} 
\right) - \Gamma \left(\displaystyle  \lambda_1 + \frac{1}{\lambda_3} \right) 
\cdot \Gamma \left(\displaystyle  \lambda_1, \left( \frac{x}{\lambda_2} 
\right)^{\lambda_3} \right)}{\lambda_3 \cdot \Gamma \left( \lambda_1 \right)}
\end{equation}

where ${\Gamma}(\cdot, \cdot)$ is the Incomplete Gamma function.

\subsection{Two-parameter Inverse Gaussian distribution}
\label{sec:ig2p}
The probability density function $p_{\rm IG}(x)$ of the two-parameter Inverse 
Gaussian distribution is

\[
p_{\rm IG}(x) =
\begin{cases}
\displaystyle
\frac{1}{\sqrt{2 \cdot \pi}} \cdot \sqrt{\displaystyle 
\frac{\lambda}{x^3}} \cdot \exp \left({\displaystyle -\frac{\lambda \left( x - 
\mu \right)^2}{2 \cdot {\mu}^2 \cdot x}} \right) & \text{if } x > 0 
\\
0 & \text{if } x \leq 0
\end{cases}
\]

where $\mu$ is the mean, and $\lambda$ is a scale parameter.

By applying the proposed method, the drift term is

\begin{equation}\label{eq:a_ig2p}
a(x) = -\alpha \cdot \left(x - \mu \right)
\end{equation}

and the diffusion term is

\begin{equation}\label{eq:b_ig2p}
b(x) = \sqrt{\frac{2 \cdot \sqrt{2 \cdot \pi} \cdot \alpha \cdot \mu \cdot \exp 
\left(\displaystyle \frac{\lambda \cdot \left( x + \mu \right)^2}{2 \cdot 
{\mu}^2 \cdot x} \right) \cdot {\rm erfc} \left( \displaystyle
\frac{\sqrt{\displaystyle \frac{\lambda}{x}} \cdot \left( x + \mu
\right)}{\sqrt{2} \cdot \mu}\right)}{\sqrt{\displaystyle 
\frac{\lambda}{x^3}}}}
\end{equation}

where ${\rm erfc}(\cdot)$ is the Complementary Error function.

\subsection{Two-parameter Lognormal distribution}
\label{sec:ln2p}
The probability density function $p_{\rm LN}(x)$ of the two-parameter Lognormal 
distribution is

\[
p_{\rm LN}(x) =
\begin{cases}
\displaystyle
\frac{1}{\sqrt{2 \cdot \pi} \cdot \sigma \cdot x} \cdot \exp 
\left({\displaystyle -\frac{\left( \log \left(x \right) - \mu \right)^2}{2 
\cdot {\sigma}^2}} \right) & \text{if } x > 0 
\\
0 & \text{if } x \leq 0
\end{cases}
\]

where $\mu$ and $\sigma$ are the mean and the standard deviation of the natural 
logarithm of variable $x$, respectively.

By applying the proposed method, the drift term is

\begin{equation}\label{eq:a_ln2p}
a(x) = -\alpha \cdot \left(x - \exp \left( \displaystyle \mu + 
\frac{\sigma^2}{2} \right) \right)
\end{equation}

and the diffusion term is

\begin{equation}
b(x) = \sqrt{b_1(x) \cdot b_2(x)}
\end{equation}

with

\begin{equation}
b_1(x) = \sqrt{2 \cdot \pi} \cdot \alpha \cdot \sigma \cdot x \cdot \exp{ 
\left(\displaystyle \mu + \frac{\sigma^2}{2} + 
\frac{\left( \log \left(x \right) - \mu \right)^2}{2 \cdot \sigma^2} \right)}
\end{equation}

and

\begin{equation}\label{eq:b_ln2p}
b_2(x) = {\rm erf} 
\left( \displaystyle \frac{\mu + \sigma^2 - \log \left( x \right)}{\sqrt{2} 
\cdot \sigma} \right) - {\rm erf} \left( \displaystyle \frac{\mu - \log \left( 
x \right)}{\sqrt{2} \cdot \sigma} \right)
\end{equation}

where ${\rm erf}(\cdot)$ is the Error function.

\subsection{One-parameter Rayleigh distribution}
\label{sec:r1p}
The probability density function $p_{\rm R}(x)$ of the one-parameter Rayleigh 
distribution is

\[
p_{\rm R}(x) =
\begin{cases}
\displaystyle
\frac{x}{\lambda^2} \cdot \exp \left(\displaystyle -\frac{x^2}{2 \cdot 
\lambda^2} \right) & 
\text{if } x > 0 
\\
0 & \text{if } x \leq 0
\end{cases}
\]

where $\lambda$ is a scale parameter.

By applying the proposed method, the drift term is

\begin{equation}\label{eq:a_r1p}
a(x) = -\alpha \cdot \left(x - \sqrt{\displaystyle \frac{\pi}{2}} \cdot \lambda 
\right)
\end{equation}

and the diffusion term is

\begin{equation}\label{eq:b_r1p}
b(x) = \sqrt{\frac{\alpha \cdot \lambda^2}{x} \cdot \left( 2 \cdot x + \sqrt{2 
\cdot \pi} \cdot \lambda \cdot \left( \exp \left(\displaystyle \frac{x^2}{2 
\cdot \lambda^2} \right) {\rm erfc} \left(\displaystyle \frac{x}{\sqrt{2} 
\cdot \lambda} \right) - 1 \right) \right)}
\end{equation}

where ${\rm erfc}(\cdot)$ is the Complementary Error function.

\subsection{Two-parameter Truncated Normal distribution}
\label{sec:tn2p}
The probability density function $p_{\rm TN}(x)$ of the two-parameter Truncated 
Normal distribution is

\[
p_{\rm TN}(x) =
\begin{cases}
\displaystyle
\sqrt{\displaystyle \frac{2}{\pi}} \cdot \frac{\exp 
\left(\displaystyle -\frac{\left( x - \mu \right)^2}{2 \cdot 
\sigma^2} \right)}{\sigma \cdot \left( 1 + {\rm erf} 
\left( \displaystyle \frac{\mu}{\sqrt{2} \cdot \sigma} \right) \right)}  & 
\text{if } x > 0 
\\
0 & \text{if } x \leq 0
\end{cases}
\]

where $\rm erf(\cdot)$ is the Error function, and $\mu$ and $\sigma$ are, 
respectively, the mean and the standard deviation of the Normal distribution 
before truncation.

By applying the proposed method, the drift term is

\begin{equation}\label{eq:a_tn2p}
a(x) = -\alpha \cdot \left(x - \mu  - \frac{\displaystyle \sigma \cdot 
\sqrt{\frac{2}{\pi}} \cdot \exp \left(\displaystyle -\frac{\mu^2}{2 
\cdot \sigma^2} \right)}{1 + {\rm erf} 
\left( \displaystyle \frac{\mu}{\sqrt{2} \cdot \sigma} \right)} \right)
\end{equation}

and the diffusion term is

\begin{equation}\label{eq:b_tn2p}
b(x) = \sqrt{2 \cdot \alpha \cdot \sigma^2 \cdot \left( 1 + \frac{\exp 
\left(\displaystyle \frac{\left( x - 2 \cdot \mu \right) \cdot x}{2 \cdot 
\sigma^2} \right) \left( {\rm erfc} \left( \displaystyle \frac{\mu - x
}{\sqrt{2} \cdot \sigma} \right) - 2 \right)}{1 + {\rm erf} 
\left( \displaystyle \frac{\mu}{\sqrt{2} \cdot \sigma} \right)} \right)}
\end{equation}

where ${\rm erfc}(\cdot)$ is the Complementary Error function.

\subsection{Two-parameter Weibull distribution}
\label{sec:w2p}
The probability density function $p_{\rm W}(x)$ of the two-parameter Weibull 
distribution is

\[
p_{\rm W}(x) =
\begin{cases}
\displaystyle
\frac{\lambda_1}{\lambda_2} \cdot \left(\displaystyle \frac{x}{\lambda_2}  
\right)^{\lambda_1-1} \cdot \exp \left( \displaystyle - \left( 
\frac{x}{\lambda_2} \right)^{\lambda_1} \right) & \text{if } x \geq 0
\\
0 & \text{if } x < 0
\end{cases}
\]

where $\lambda_1$ is a shape parameter and $\lambda_2$ is a scale parameter.

By applying the proposed method, the drift term is

\begin{equation}\label{eq:a_w2p}
a(x) = -\alpha \cdot \left( x - \lambda_2 \cdot \Gamma \left( 1 + 
\frac{1}{\lambda_1} \right) \right)
\end{equation}

and the diffusion term is

\begin{equation}
b(x) = \sqrt{b_1(x) \cdot b_2(x)}
\end{equation}

with

\begin{equation}
b_1(x) = 2 \cdot \alpha \cdot \frac{\lambda_2}{{\lambda_1}^2} \cdot x \cdot
\left( \frac{\lambda_2}{x} \right)^{\lambda_1}
\end{equation}

and

\begin{equation}\label{eq:b_w2p}
b_2(x) = \lambda_1 \cdot \exp 
\left( \left(\frac{x}{\lambda_2} \right)^{\lambda_1} \right) \cdot \Gamma 
\left( 1 + \frac{1}{\lambda_1}, \left(\frac{x}{\lambda_2} \right)^{\lambda_1} 
\right) - \Gamma \left( \frac{1}{\lambda_1} \right)
\end{equation}

where $\Gamma(\cdot)$ is the Gamma function, and $\Gamma(\cdot, \cdot)$ is the 
Incomplete Gamma function.

\section{Conclusions} \label{sec:con}
In this paper, we develop a systematic method to construct wind speed models 
based on stochastic differential equations. We apply a novel, analytically exact 
approach to define the formulation of the drift and diffusion terms of a 
stochastic differential equation in order to reproduce the given stationary 
probability distribution and exponential autocorrelation characterizing the wind 
speed. This new approach accurately reproduces both the probability 
distribution 
and the autocorrelation of the wind speed, as opposed to existing methods that 
are approximated. The application of the proposed method is straightforward 
and 
can be carried out systematically. Proof of that is the collection of models 
developed in the paper for different probability distributions proposed in the 
literature to describe the wind speed behaviour. Finally, the proposed method 
is general and can be applied to model any stationary process with exponential 
autocorrelation. Future work will focus on the definition of SDE-based models 
for processes with autocorrelation other than exponential. 

\bibliographystyle{plain}
\bibliography{references_sde}

\end{document}